\documentstyle[preprint,aps,epsf]{revtex}
\newcommand{\beq}{\begin{equation}}
\newcommand{\eeq}{\end{equation}}
\newcommand{\ds}{\displaystyle}
\newcommand{\beqar}{\begin{eqnarray}}
\newcommand{\eeqar}{\end{eqnarray}}

\begin{document}
\tightenlines
\draft

\title{
 Equation of state of resonance-rich matter in the central cell in 
 heavy-ion collisions at $\sqrt{s}=200${\it A\/} GeV
}
\author{
L.V.~Bravina,$^{1,2}$ E.E.~Zabrodin,$^{1,2}$ 
S.A.~Bass,$^{3}$ M.~Bleicher,$^{4}$ M.~Brandstetter,$^{5}$ 
Amand~Faessler,$^{1}$ C.~Fuchs,$^{1}$, W.~Greiner,$^{5}$
S.~Soff,$^{6}$ and H.~St{\"o}cker$^{5}$ 
}
\address{
$^1$Institute for Theoretical Physics, University of T\"ubingen,\\
D-72076 T\"ubingen, Germany \\
$^2$Institute for Nuclear Physics, Moscow State University,
RU-119899 Moscow, Russia \\
$^3$National Superconducting Cyclotron Laboratory, Michigan State 
University, East Lansing, MI 48824-1321, USA \\
$^4$Nuclear Science Division, Lawrence Berkeley Laboratory, 
Berkeley, CA 94720, USA \\
$^5$Institute for Theoretical Physics, University of Frankfurt,\\
D-60325 Frankfurt, Germany \\ 
$^6$Gesellschaft f\"ur Schwerionenforschung, D-64220 Darmstadt, 
Germany \\ 
}

\date{\today}
\maketitle

\begin{abstract}
The equilibration of hot and dense nuclear matter produced in the 
central cell of central Au+Au collisions at RHIC 
($\sqrt{s}=200${\it A} GeV) energies is studied within a microscopic 
transport model.
The pressure in the cell becomes isotropic at 
$t\approx 5$ fm/$c$ after beginning of the collision. Within the next
15 fm/$c$ the expansion of matter in the cell proceeds almost
isentropically with the entropy per baryon ratio $S/A \cong 150$,
and the equation of state in the $(P,\varepsilon)$ plane has a 
very simple form, $P=0.15\varepsilon$.
Comparison with the statistical model of an ideal hadron gas
indicates that the time $t \approx 20$ fm/$c$ may be too short
to reach the fully equilibrated state. Particularly, the creation of 
long-lived resonance-rich matter in the cell decelerates the 
relaxation to chemical equilibrium. This resonance-abundant state can 
be detected experimentally after the thermal freeze-out of particles. 
\end{abstract}


\widetext 

\section{Introduction}
\label{sec1}

The assumption that strongly interacting hadronic (or rather partonic)
matter, produced in nucleus-nucleus collisions at high energy, can 
reach a state of local equilibrium (LE) \cite{Fer50,Land53,Bjor83} is 
one of the most important topics in relativistic heavy-ion program
\cite{Tor99}.
The degree of equilibration can be checked by fitting the measured
particle yields and transverse momentum spectra to that of the thermal 
model in order to extract the conditions of the fireball at the 
chemical and thermal freeze-out (see, e.g., 
\cite{BrSt96,Bec,Raf,ClRe99,CEST97,UH99,Soll97,YeGo99} 
and references therein). Here the equilibrium particle abundances,
which correspond to a certain temperature, $T$, baryon chemical
potential, $\mu_{\rm B}$, and strangeness chemical potential, 
$\mu_{\rm S}$, can be determined. However, the analysis is 
complicated, e.g., by the presence of collective flow of particles 
and the non-homogeneity of the baryon charge distribution in the 
reaction volume. The volume of the fireball is also taken as a free 
parameter in the thermal model. Various non-equilibrium microscopic 
transport-, string-, partonic-, etc. models have been applied to 
verify the appearance of at least local equilibrium in the course of 
heavy-ion collisions at relativistic and ultrarelativistic energies 
\cite{GeKa93,Cass,Faes94,Bass98,Blei98,SHSX99,CBJ99,lv98plb,lv99prc,
urqmd,Belk98,Brat00}.
To reduce the number of unknown parameters and simplify the analysis 
it has been suggested \cite{lv98plb,lv99prc} to 
examine the equilibrium conditions in the central cell of relativistic 
heavy-ion collisions, simulated by one of the microscopic models.
For these purposes we employ the microscopic Ultra-relativistic 
Molecular Dynamics (UrQMD) transport model, which nicely describe the 
available experimental data on hadron-hadron and nucleus-nucleus 
collisions in a broad energy range \cite{ urqmd}.

The central cell of symmetric heavy-ion collisions is a convenient 
system because the velocity of its center of mass is essentially 
zero. The cell should be neither too small nor too large. Previous
studies at energies from 10.7{\it A\/} GeV (AGS) to 160{\it A\/} GeV 
(SPS) \cite{lv98plb,lv99prc} have shown that the
cubic cell with volume $V = 5\times 5\times 5 = 125$ fm$^3$ is well
suited for the analysis. 
The aim of the present paper is to study the relaxation of hot nuclear
matter, simulated within the microscopic model, in the central cell in
Au+Au interactions at RHIC ($\sqrt{s}=200${\it A\/} GeV) energy.

The paper is organized as follows. The conditions of kinetic 
equilibrium in the cell are discussed in Sec.~\ref{sec2}.
Section~\ref{sec3} describes the statistical model (SM) of an ideal
hadron gas employed for the comparison with the microscopic
calculations. The relaxation of hot nuclear matter in the cell
to thermal and chemical equilibrium is studied in Sec.~\ref{sec4}.
Finally, conclusions are drawn in Sec.~\ref{sec5}.

\section{Kinetic equilibrium}
\label{sec2}

\begin{figure}[htb]
\centerline{\epsfysize=18cm \epsfbox{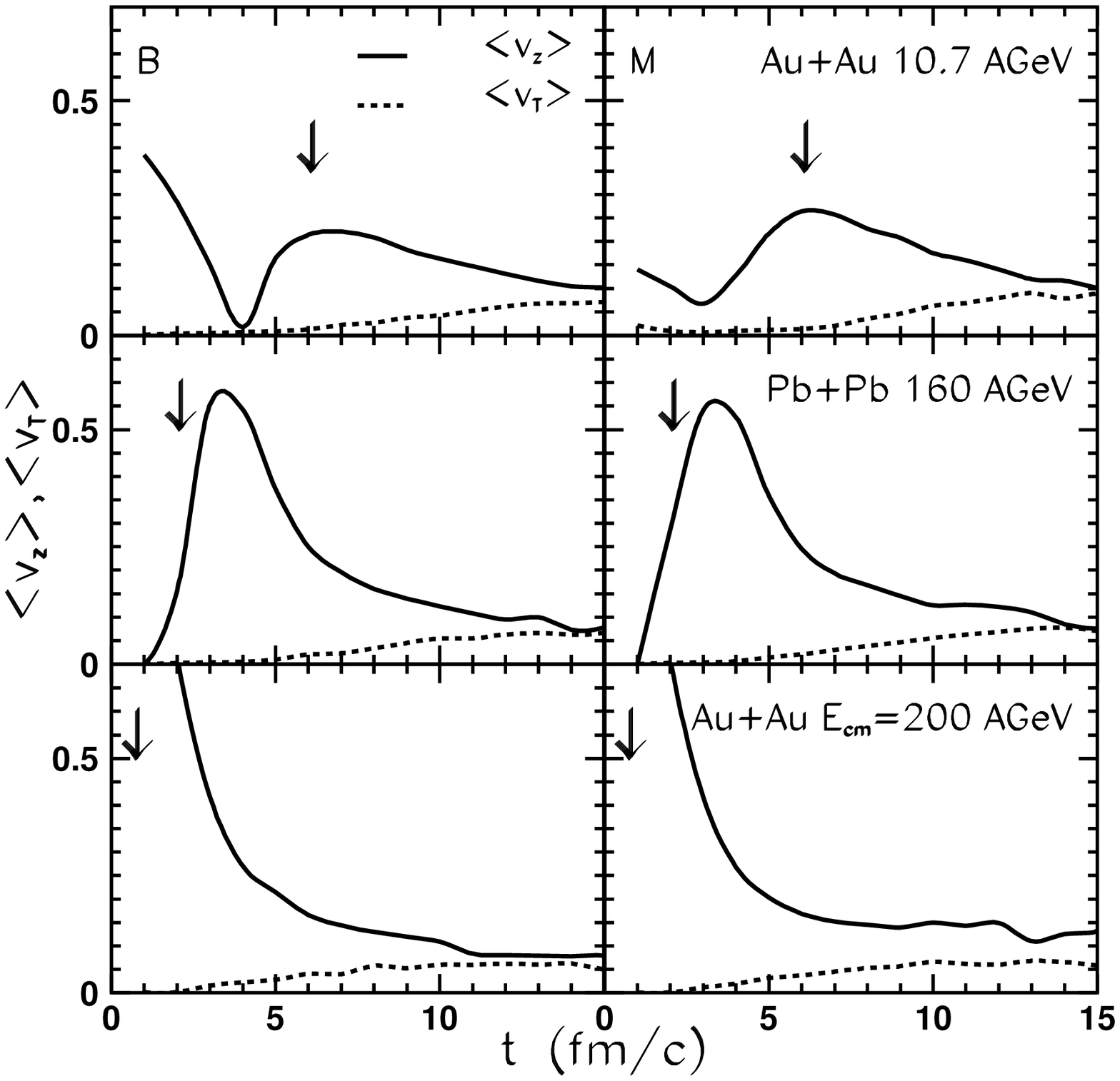}}
\caption{
Time evolution of the
velocities of longitudinal (solid lines) and transverse (dashed
lines) flows of baryons (left panels) and mesons (right panels)
in an asymmetric cell $0 \leq \{x,y,z\} \leq 2.5$ fm of central
heavy-ion collisions at $E_{lab} = 10.7${\it A\/}GeV (upper row),
$E_{lab} = 160${\it A\/}GeV (middle row), and $\sqrt{s} = 
200${\it A\/} GeV (bottom row), respectively. Arrows indicate the 
times $t_{cr}$ needed for Lorentz-contracted nuclei to pass 
through each other.
}
\label{fig1}
\end{figure}

First, the kinetic equilibrium has to be verified.
The collective flow in the cell should be isotropic and small, so it
cannot significantly distort the momentum distributions of particles.
As shown in Fig.~\ref{fig1}, the average longitudinal velocity in the 
central cell at RHIC energies reaches its maximum at 
$\approx 1$ fm/$c$ after the Lorentz-contracted nuclei have completely 
passed through each other. The longitudinal flow rapidly drops and 
converges to the developing transverse flow. The longitudinal 
collective velocity of particles becomes smaller than $v_{flow} = 
0.15\, c$ at $t = 5-6$ fm/$c$ at RHIC, while the similar process 
takes $t = 8-10$ fm/$c$ at the AGS and SPS energies. This gives a 
small correction, only $\langle m_N\, v_{flow}^2/2 \rangle \approx 7$ 
MeV for the nucleon spectra, and less than a MeV for pions, whereas 
the characteristic temperatures are around 140 MeV. 

Velocity distributions of hadron species become isotropic and nearly
Maxwellian quite soon \cite{lv98plb,lv99prc}. Isotropy of the velocity 
distributions results in pressure isotropy. The pressure in the 
longitudinal direction in the cell, calculated according to virial 
theorem \cite{Bere92}, is compared in Fig.~\ref{fig2}(a) with the 
transverse pressure. The time of convergence of longitudinal pressure 
to the transverse one decreases from 10 fm/$c$ to 5 fm/$c$ with rising 
incident energy from AGS to RHIC, respectively.
Thus, the kinetic equilibrium in the central cell in Au+Au collisions 
at RHIC energy is reached at $t \approx 5$ fm/$c$.

\section{Statistical model of an ideal hadron gas}
\label{sec3}

To verify that the matter in the cell is in thermal and chemical
equilibrium one has to compare the snapshot of hadron yields and 
energy spectra in the cell with the equilibrium spectra of hadrons
obtained in the statistical model (SM) of an ideal hadron gas 
containing essentially the same number of baryonic and mesonic 
degrees of freedom \cite{Belk98}. We recall the procedure briefly:
the values of energy density, $\varepsilon$, baryon density, 
$\rho_{\rm B}$, and strangeness density, $\rho_{\rm S}$, extracted   
from the microscopic calculations in the cell, are used as an input
to the system of nonlinear equations of the SM
\beqar
\ds
\label{eq1}
\varepsilon^{\rm mic} &=& \frac{1}{V}\sum_i
            E_i^{\rm SM}(T,\mu_{\rm B},\mu_{\rm S}), \\
\label{eq2}
\rho_{\rm B}^{\rm mic}&=& \frac{1}{V}\sum_i
            B_i\cdot N_i^{\rm SM}(T,\mu_{\rm B},\mu_{\rm S}), \\
\label{eq3}
\rho_{\rm S}^{\rm mic}&=& \frac{1}{V}\sum_i
            S_i\cdot N_i^{\rm SM}(T,\mu_{\rm B},\mu_{\rm S}),
\eeqar
containing the  baryon charge, $B_i$, and strangeness, $S_i$, of
the hadron species $i$. This procedure enables one to determine
temperature, $T$, baryon chemical potential, $\mu_{\rm B}$, and
strangeness chemical potential, $\mu_{\rm S}$. If the set of three
parameters $(T, \mu_{\rm B}, \mu_{\rm S})$ is fixed, all
macroscopic characteristics of the equilibrated system can be
determined unambiguously. For instance,
the particle yields,  $N_i^{\rm SM}$, total energy,
$E_i^{\rm SM}$, and pressure, $P^{\rm SM}$, are calculated within 
the SM via the Gibbs distribution function 
\beq
\ds
f_i(p,m_i) \propto \exp{\left( - \sqrt{p^2 + m_i^2}/T + \mu_i/T 
\right) } , 
\label{eq4}
\eeq
where $p$ and $m_i$ are the momentum and the mass of the hadron 
species $i$, respectively. Namely,
\beqar
\ds
N_i^{\rm SM} &=& \frac{V g_i}{(2\pi\hbar)^3}\int f(p,m_i)\, d^3 p ,\\
\label{eq5}
E_i^{\rm SM} &=& \frac{V g_i}{(2\pi\hbar)^3}\int (p^2+m_i^2)^{1/2}\,
f(p,m_i)\, d^3 p , \\
\label{eq6}
P^{\rm SM} &=& \sum_i \frac{g_i}{(2\pi\hbar)^3}\int 
\frac{p^2}{3 (p^2+m_i^2)^{1/2}}\, f(p,m_i)\, d^3 p ,
\label{eq7}
\eeqar
with $g_i$ being the degeneracy factor.
The entropy density $s = S/V$ can be determined either 
via the distribution function (\ref{eq4})
\beq
\ds
s = -\sum_i \frac{g_i}{\left( 2\pi\hbar \right)^3} \int_0^{\infty}
   f(p,m_i)\, \left[ \ln{f(p,m_i)}-1 \right] \, d^3 p ,
\label{eq8}
\eeq
or directly from Gibbs thermodynamical identity
\beq
\ds
T s =  \varepsilon + P^{\rm SM} - \mu_{\rm B} \rho_{\rm B} -
\mu_{\rm S} \rho_{\rm S} .
\label{eq9}
\eeq
The chemical potential $\mu_i$ of the $i$th hadron is determined by 
its baryon and strangeness charge, 
$\mu_i = \mu_{\rm B} B_i + \mu_{\rm S} S_i$. Thermal and chemical
equilibrium is assumed to 
occur in the cell when the spectra of hadrons in the microscopic 
cell calculations become close to the spectra predicted by the SM. 

\section{Relaxation to chemical and thermal equilibrium}
\label{sec4}

\begin{figure}[htb]
\centerline{\epsfysize=18cm \epsfbox{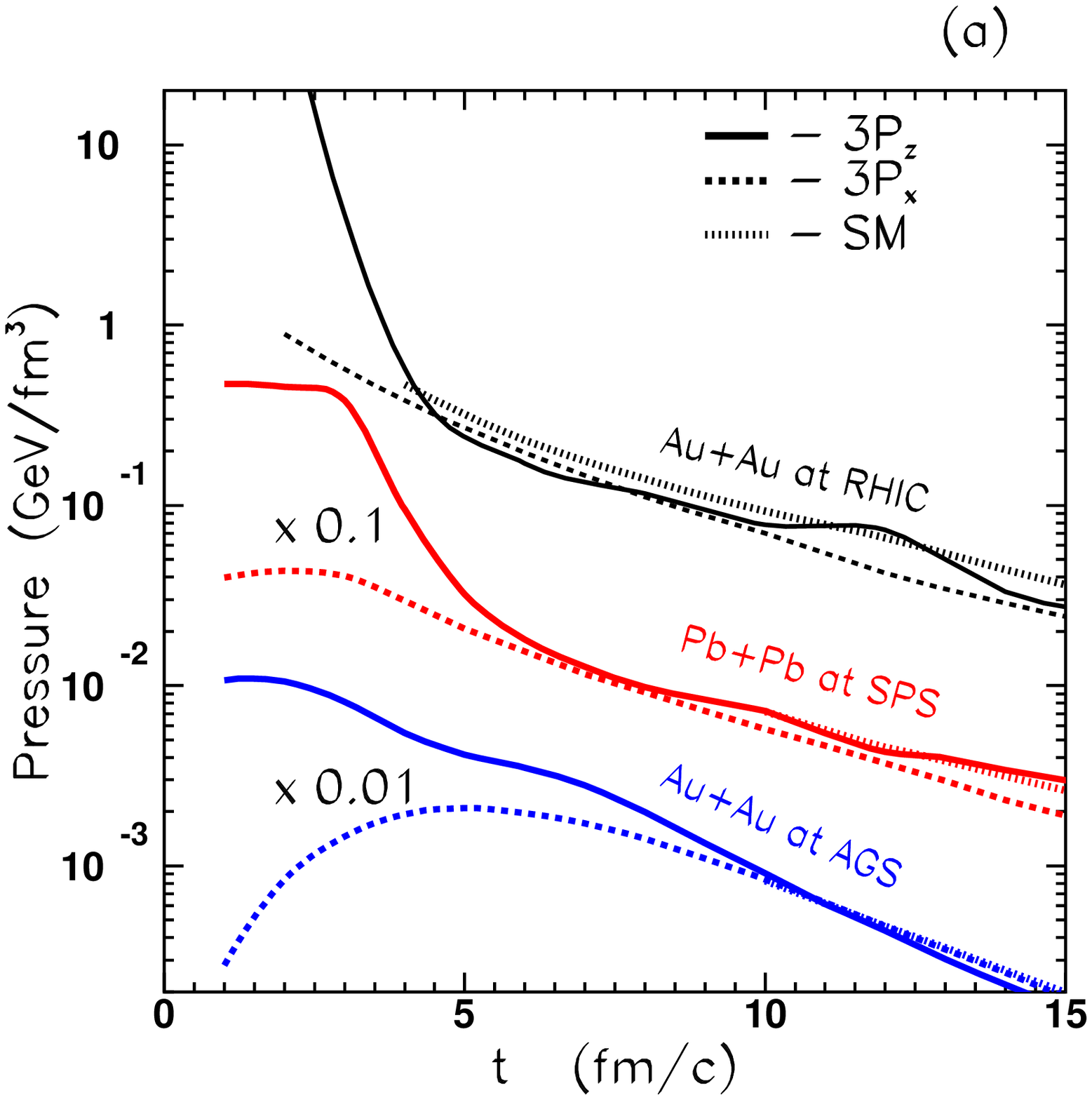}}
\caption{
(a) The longitudinal (solid lines) and the transverse 
(dashed lines) diagonal components of the pressure tensor $P$ in 
the central cell of heavy-ion collisions at AGS, SPS, and RHIC
energies compared to the SM results (dotted lines).\\
(b) Time evolution of the entropy per baryon ratio, $s/\rho_{\rm_B} 
= S/A$, in the central cell with $V = 125$ fm$^3$ in heavy-ion 
collisions at AGS, SPS, and RHIC energies, respectively.
Solid lines denote microscopic calculations with the UrQMD model,
dashed lines show the predictions of the statistical model.
}
\centerline{\epsfysize=18cm \epsfbox{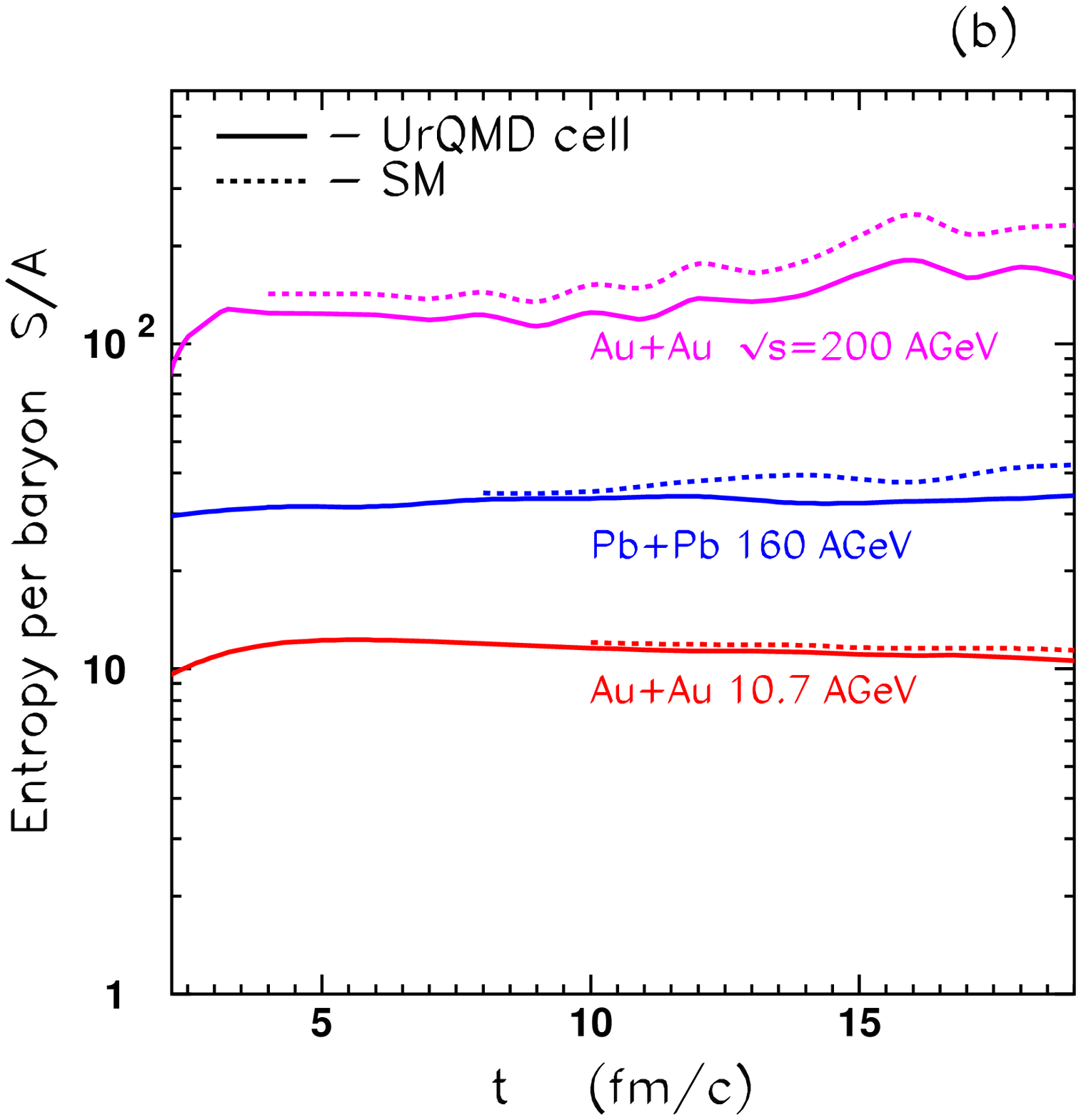}}
\label{fig2}
\end{figure}

At the isotropic stage the total microscopic pressure is close to
the grand canonical pressure, $P^{\rm SM}$, as shown in 
Fig.~\ref{fig2}(a). Both pressures converge at about $t=5$ fm/$c$,
which is chosen as a starting point for the comparison with the SM. 
The input, $\{ \varepsilon, \rho_{\rm B}, \rho_{\rm S} \}$, and 
output, $\{ T, \mu_{\rm B}, \mu_{\rm S} \}$, parameters are listed 
in Table~\ref{tab1} together with the microscopic pressure, entropy
density, and entropy per baryon in the cell. The final time of the 
calculation, defined from conventional freeze-out conditions
$\varepsilon \approx 0.1$ GeV/fm$^3$ or $\rho_{tot} \approx 0.5 
\rho_0$ \cite{lv99prc}, corresponds to $t = 20-21$ fm/$c$. One may 
see that in spite of different initial conditions the freeze-out
time in the central cell in heavy-ion collisions at RHIC energies is 
similar to corresponding freeze-out times at AGS \cite{lv98plb} and 
SPS \cite{lv99prc} energies. In accordance with general estimates
\cite{ClRe99,BMS98} the baryochemical potential at RHIC energies is
small while the temperatures are well above the anticipated 
temperature for the QCD phase transition, $T \approx 160 \pm 10$ MeV.
The entropy per baryon in the cell varies slightly after the beginning 
of the kinetic equilibrium stage. A comparison with 
$s/\rho_{\rm B}$ at lower energies is shown in Fig.~\ref{fig2}(b). 
The hadron-string matter in the central cell seems to expand 
isentropically with $s/\rho_{\rm B} \equiv S/A \approx 12$ (AGS), 
32 (SPS), and 150 (RHIC). Note that the results of the simulations
at AGS and SPS energies are intriguingly close to the entropy per
baryon values extracted from the thermal model fit to experimental
data, namely, $(S/A)^{AGS} \approx 14$ and $(S/A)^{SPS} \approx 36$
\cite{ClRe99}. It is most interesting to compare the predicted value 
$(s/\rho_{\rm B})^{\rm RHIC} = 150 - 170$ to the upcoming RHIC data.
Together, isotropy in the pressure sector, isotropic and nearly 
Maxwellian velocity distributions of hadrons, and almost isentropic 
expansion of matter in the central cell strongly support the idea 
that a hydrodynamic regime is established after a certain time. In 
the following, we extract the equation of state (EOS) of 
hadron-resonance-string matter
from the microscopic simulations at the quasi-equilibrium stage of
nuclear collisions, which is part of this hydrodynamic picture. 

\begin{figure}[htb]
\centerline{\epsfysize=18cm \epsfbox{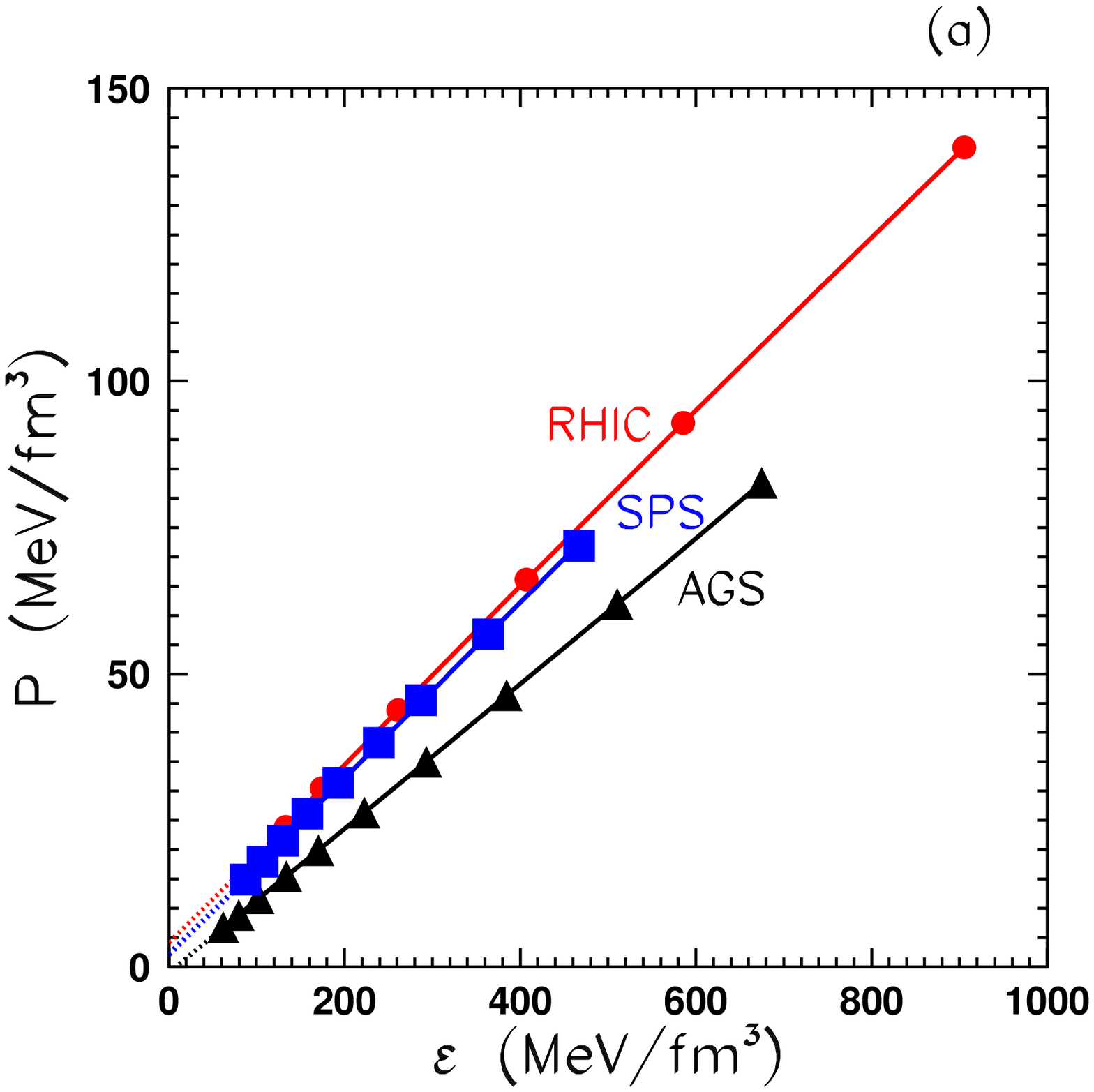}}
\caption{
(a) The evolution of pressure $P$ and baryon density 
$\varepsilon$ in the central cells of the heavy-ion collisions
at AGS, SPS, and RHIC energies.\\
(b) The same as (a) but for the $(T, \mu_{\rm B})$-plane.
Solid symbols correspond to the stage of kinetic equilibrium,
open symbols indicate the preequilibrium stage.
The hatched area shows the expected region of the quark-hadron
phase transition.
}
\centerline{\epsfysize=18cm \epsfbox{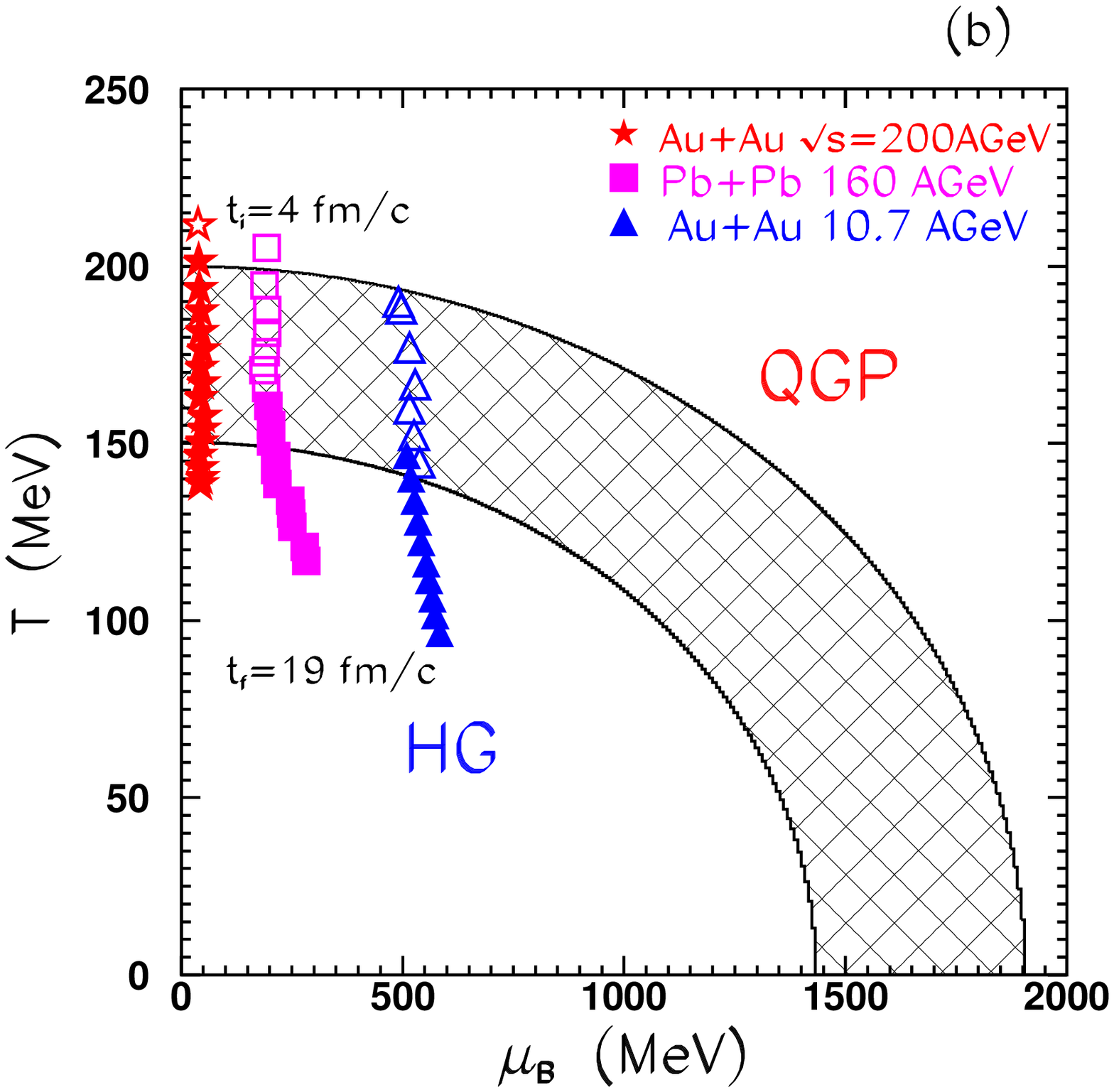}}
\label{fig3}
\end{figure}

The evolution of the pressure with the energy 
density is depicted in Fig.~\ref{fig3}(a) for AGS, SPS, and RHIC 
energies in the central cell, respectively. Several interesting 
facts can be gained from this figure. First, the pressure drops 
linearly with the decreasing energy density for all three energies. 
It means that the ratio $P /\varepsilon$ remains constant for the 
whole time interval of the (quasi)equilibrium stage. Thus the 
equation of state in the $(P,\varepsilon)$ plane has a very simple 
form: $P = 0.12\, \varepsilon$ at AGS, and $P = 0.15\, \varepsilon$ 
at SPS and RHIC. We see that despite the significant difference in 
the center-of-mass energy of the nuclei, colliding at $\sqrt{s} =
17${\it A\/} GeV vs. $\sqrt{s} = 200${\it A\/} GeV, the ratio
$P /\varepsilon$ in the central cell is saturated already at 
SPS energies.

Figure~\ref{fig3}(b) presents the evolution of the EOS in the 
$(T, \mu_B)$ plane. The baryon chemical potential in the cell
increases with decreasing temperature at AGS and SPS energies,
but remains almost constant (and small) at RHIC. Also, all
three curves can be well approximated by a simple linear
dependence. This fact deserves explanation. Substituting $P = a 
\varepsilon$ in Eq.~(\ref{eq9}) and omitting the term
$\mu_{\rm S} \rho_{\rm S}$, because the net strangeness density
in the cell is quite small, we get
\beq
\ds
T \frac{s}{\rho_{\rm B}} = \frac{\varepsilon (1+a)}{\rho_{\rm B}}
 - \mu_{\rm B}\ .
\label{eq10}
\eeq
If the ratio ${\varepsilon / \rho_{\rm B}}$ is constant,
then the temperature $T$ would depend linearly only on the baryon 
chemical potential $\mu_{\rm B}$, and vice versa. However, the
energy per baryon slightly decreases within the considered time
interval for all three energies. At the final stage it drops to
approximately 80-90\% of its initial value. Therefore, from
Fig.~\ref{fig3}(b) it follows that the evolution of temperature 
as a function of ${\varepsilon / \rho_{\rm B}}$ and $\mu_{\rm B}$
is mainly determined by the change of energy per baryon, and not
the baryon chemical potential.  

\begin{figure}[htb]
\centerline{\epsfysize=18cm \epsfbox{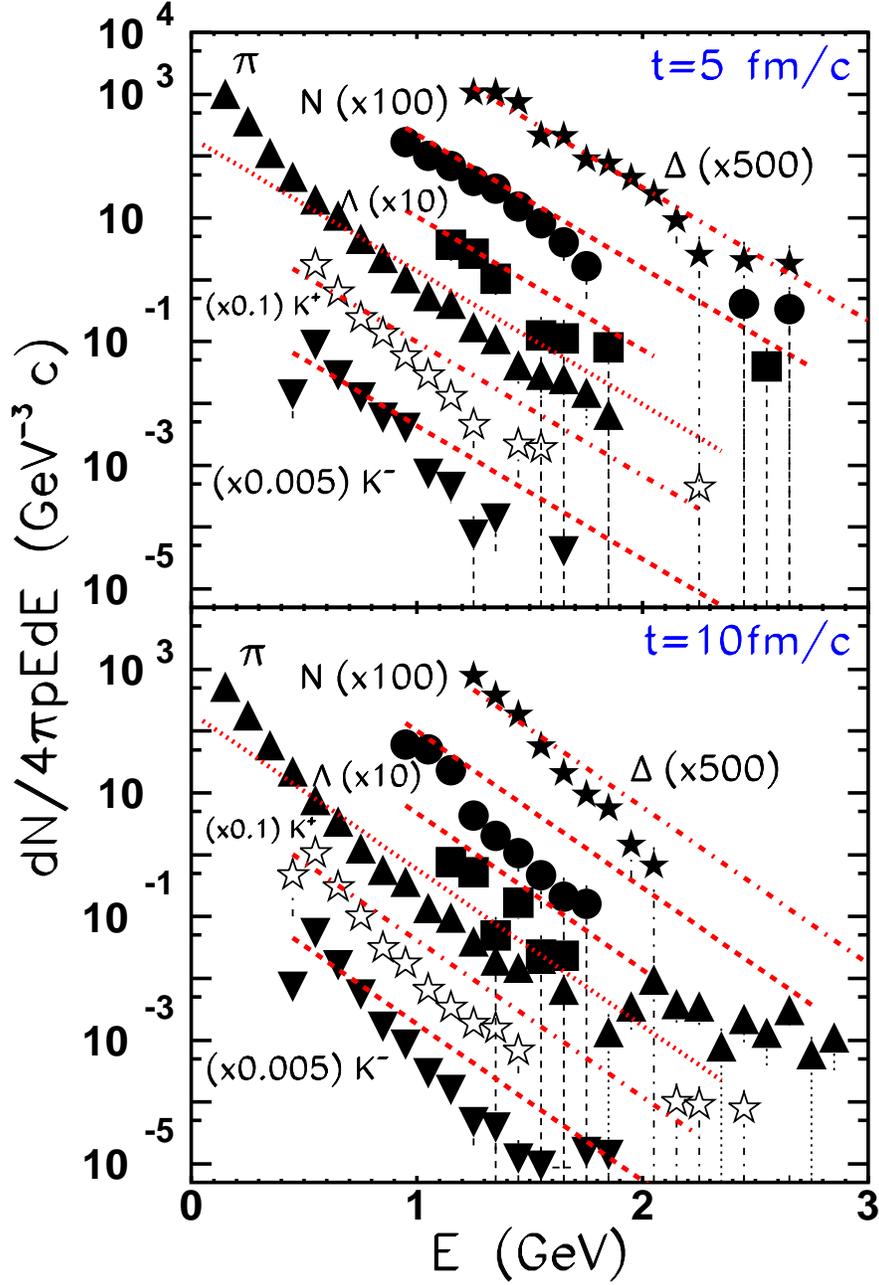}}
\caption{
Energy spectra of $N$ (circles), $\Lambda $ (squares),
$\pi $ (triangles up), $K^+$ (open stars), $K^-$ (triangles down) 
and $\Delta$ (solid stars) in the central cell of 
Au+Au collisions at RHIC at $t=5$ fm/$c$ (upper panel) and 
$t=10$ fm/$c$ (bottom panel). Dashed lines show the predictions
of the SM.
}
\label{fig4}
\end{figure}

Energy spectra, $d N/(4\pi pEdE)$, of $\pi$'s, $N$'s, $\Lambda$'s,
$\Delta$'s and $K$'s in the cell at the stage of kinetic
equilibrium are shown in Fig.~\ref{fig4} at two different times,
$t = 5$ fm/$c$ and $t = 10$ fm/$c$ at RHIC energies. 
At $t = 5$ fm/$c$ the spectra of baryons seem to be in a reasonable 
agreement with those of the SM, while the slopes
of the meson spectra are steeper compared to the predictions of 
the statistical model. This means, that the apparent temperatures
of mesons, especially pions, are lower than the temperature
given by the SM. Moreover, the energy spectra of pions can be 
decomposed on two components representing low and high energy pions.
At $t = 5$ fm/$c$ the fit by two exponents to the pion spectrum 
yields the temperatures $T_{low}^\pi \approx 114$ MeV and 
$T_{high}^\pi \approx 158$ MeV, while at $t = 10$ fm/$c$ the results
of the fit are $T_{low}^\pi \approx 100$ MeV and $T_{high}^\pi 
\approx 150$ MeV, respectively. Although the apparent temperature of 
pions from the high energy tail of the energy spectrum is closer to the
temperature given by the SM, $T^{SM}_{t=10} = 171$ MeV, one may
conclude that the thermal equilibrium in the central cell is not 
reached yet. The slopes of baryon spectra in the microscopic
calculations at $t = 10$ fm/$c$ are also steeper than the SM slope.
Since the energy density $\varepsilon$ is the same in both models,
the lower temperatures of hadronic spectra in the microscopic case
indicate that the hot matter in the cell is not in chemical
equilibrium. Therefore, one might expect that the fractions of
mesons and resonances in the UrQMD cell are overpopulated. These 
extra-particles consume significant part of the total energy and
effectively ''cool" the hadron cocktail. The reheating proceeds
via the absorption of mesons in the processes like $\pi \pi
\rightarrow \rho,\ \rho \rho \rightarrow \pi \pi$, or $\pi N
\rightarrow \Delta,\ \Delta N \rightarrow N N$, etc. Thus, our
next step is to study the time evolution of hadron abundances in
the cell. Note also that the difference between the temperatures 
of meson and baryon spectra can be explained by the
fact that baryons experience many more elastic collisions and
interactions with the resonance production per particle, which
drive the system toward thermal equilibrium, than mesons. The mean
number of interactions per, e.g., nucleon or delta increases
from 5 interactions at $t = 5$ fm/$c$ to 10 interactions at 
$t = 10$ fm/$c$. In contrast,
pions (kaons), which are readily absorbed and produced by the
resonance-string hadron matter, suffer only 0.5 (1) elastic
collision per particle at $t = 5$ fm/$c$ and 1.8 (2.5) ones
at $t = 10$ fm/$c$.

\begin{figure}[htb]
\centerline{\epsfysize=18cm \epsfbox{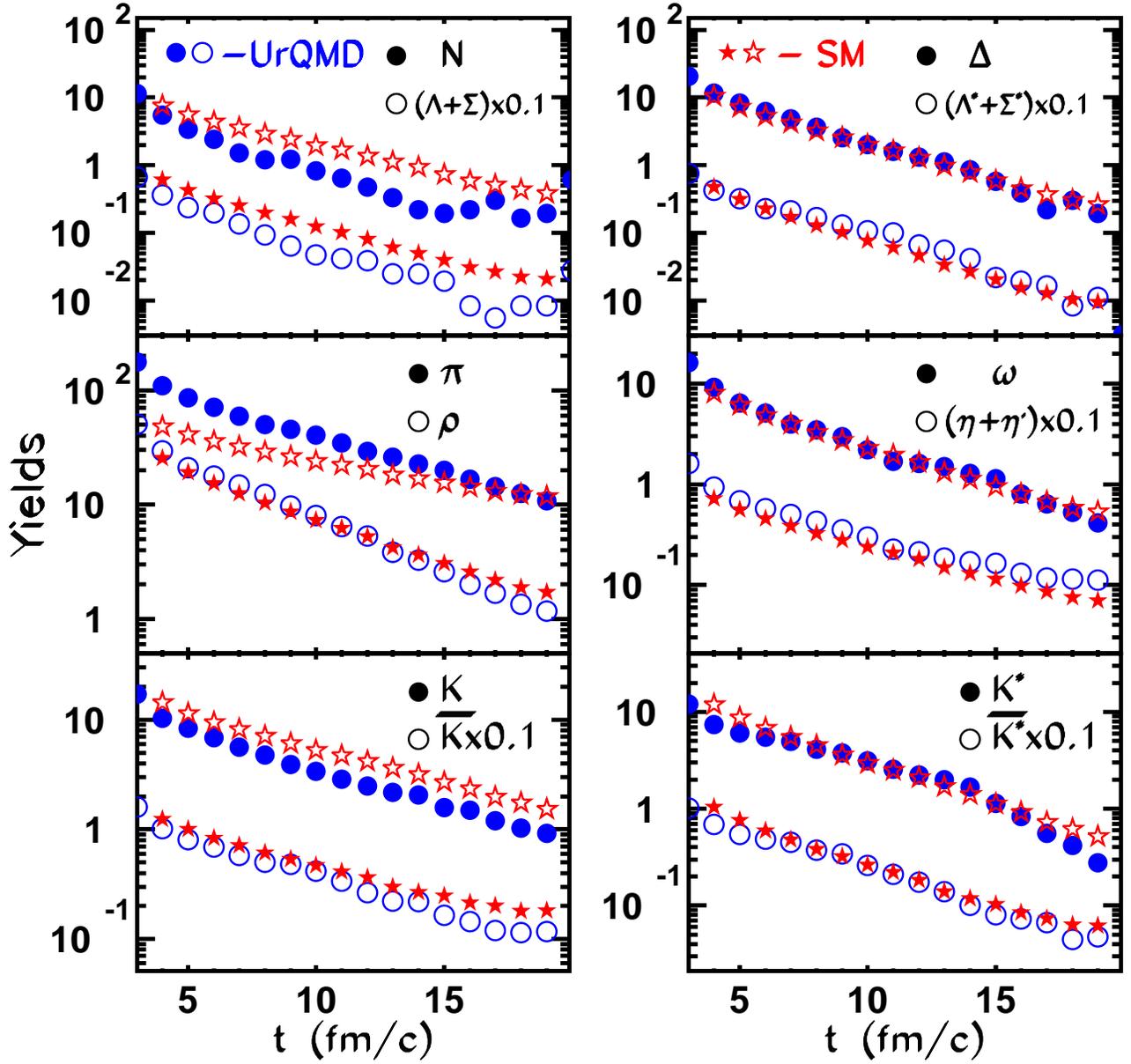}}
\caption{
The yields of main hadron species in the central cell of Au+Au 
collisions at $\sqrt{s} = 200${\it A\/} GeV as a function of time as 
obtained in the model UrQMD (circles) together with the 
predictions of the SM (stars).
}
\label{fig5}
\end{figure}

Figure \ref{fig5} depicts the yields of the main hadron species
in the cell within the time interval 5 fm/$c$ $\leq t \leq$ 19
fm/$c$. It is interesting that microscopic spectra of pions, which 
are underestimated by the SM in the central cell at lower energies 
\cite{lv99prc}, converge to the SM predictions at $t \approx 15$ 
fm/$c$. Also, the statistical model overestimates yields of 
nucleons, lambdas, and kaons, while the yields of both baryon and 
meson resonances are reproduced quite well. Since the baryon number 
and the strangeness are conserved in strong interactions, where is
the rest of the hypercharge, $Y = {\rm B + S}$, in the UrQMD 
calculations in the cell? As seen from Table~\ref{tab2}, where the 
partial densities of baryons $R_{\rm B}$ and antibaryons
$R_{\rm \bar{B}}$ in the microscopic and
macroscopic model are listed (the net values
$\rho_{\rm B} \equiv R_{\rm B} - R_{\rm \bar{B}}$ are listed in
Table~\ref{tab1}), the
total yields of baryons and antibaryons in the SM are larger than 
those of the UrQMD. The hadron-resonance-string matter in the cell 
is not in chemical equilibrium; that is why the density of
antibaryons is 2-3 times smaller than the equilibrium values.
The results on strangeness densities are listed in Table~\ref{tab3}
for the strange baryons, mesons, and their antiparticles, 
respectively. Here the net strangeness is $\rho_{\rm S} 
\equiv R_{\rm S}^{\rm B} + R_{\rm \bar{S}}^{\rm \bar{B}} +
R_{\rm S}^{\rm M} +  R_{\rm \bar{S}}^{\rm M}$.
In line with the previous observation, the SM predicts
significantly larger abundances of both strange baryons and strange
antibaryons at $5 \leq t \leq 11$ fm/$c$ in the cell, while the
densities of strange mesons are pretty similar. It is worth noting
here that strangeness is underpredicted by the UrQMD if compared to
experimental data at SPS energies, see \cite{Soff99}. 
At $t\approx 12$ fm/$c$
the situation turns around: microscopic yields of strange baryons 
become closer to those of the SM, but the predictions for strange
mesons diverge. Note also that up to $t = 13$ fm/$c$ the net
strangeness of baryons in microscopic calculations is larger
compared to the net strangeness of baryons in the SM at the expense 
of the net mesonic strangeness. After $t = 13$ fm/$c$ the fractions 
of the net strangeness deposited in baryonic and mesonic sectors in 
the microscopic calculations coincide with the SM results.

\begin{figure}[htb]
\centerline{\epsfysize=18cm \epsfbox{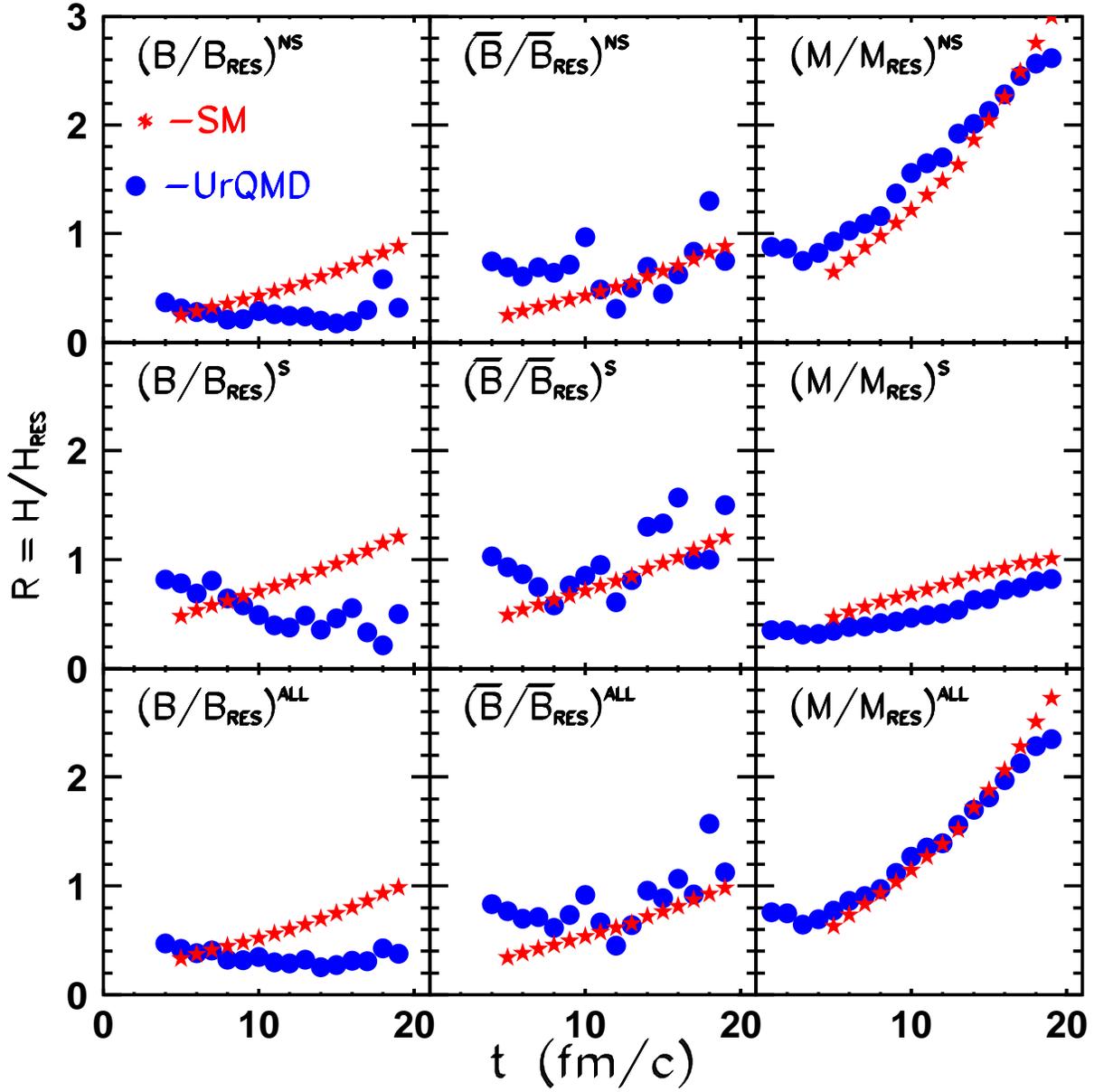}}
\caption{
Time evolution of the hadron to resonance ratio $R = H/H_{res}$ in
the central cell of Au+Au collisions at RHIC shown separately for
baryons (left panels), antibaryons (middle panels), and mesons
(right panels), as well as for non-strange hadrons (upper row),
strange hadrons (middle row), and total hadron yields (bottom row).
Circles denote the UrQMD predictions, stars correspond to the SM
results.
}
\label{fig6}
\end{figure}

Finally, the ratios of hadronic abundances are studied (see 
Fig.~\ref{fig6}). Here the results are presented separately for 
non-strange and strange baryons and mesons. In the baryon sector
the resonances dominate over the strange and non-strange baryons 
until the end of the simulations. This can be taken as an indication 
of the creation of long-lived resonance-rich matter. Recall that the 
formation of resonance abundant matter plays an important role in
the evolution of the system produced in nuclear collisions at lower
bombarding energies \cite{Ho95}. It has been suggested that the
significance of the resonances, mainly $\Delta$'s, for the system
development should be diminished with rising center-of-mass energy
of the collisions. This assumption is not confirmed in our 
simulations. The fraction of baryon resonances is almost 70\% of
all baryons in the cell at RHIC at $5 \leq t \leq 19$ fm/$c$,
while at SPS and AGS the number of baryon resonances decreases from
70\% to 35\%, and from 60\% to 25\%, respectively. The meson 
fractions of resonances shrink within the time interval 
$5 \leq t \leq 19$ fm/$c$ from 60\% to 30\% (RHIC), 50\% to 20\%
(SPS), and 40\% to 15\% (AGS). But at RHIC energies the hot hadronic 
matter in the cell as well as in the whole volume of the reaction
is meson dominated. The mesons, baryons, and antibaryons carry
90\%, 7\%, and 3\% of the total number of particles in the RHIC
cell at $t \geq 10$ fm/$c$ (cf. 85\%, 14.5\%, 0.5\% at SPS and
50\%, 50\%, 0\% at AGS). The microscopic ratios for mesons 
(Fig.~\ref{fig6}, right panels) seem to be very close to the SM
ratios. However, one has to keep in mind that the temperatures
given by the SM fit are 40-50 MeV higher than the apparent
temperatures of meson species. Since the freeze-out occurs at
$t \approx 21$ fm/$c$ in the central cell at RHIC energies, the
matter in the cell is frozen before reaching thermal and chemical
equilibrium. This circumstance significantly complicates the
extraction of the chemical and thermal freeze-out parameters by
means of the standard thermal model fit.

\begin{figure}[htb]
\centerline{\epsfysize=18cm \epsfbox{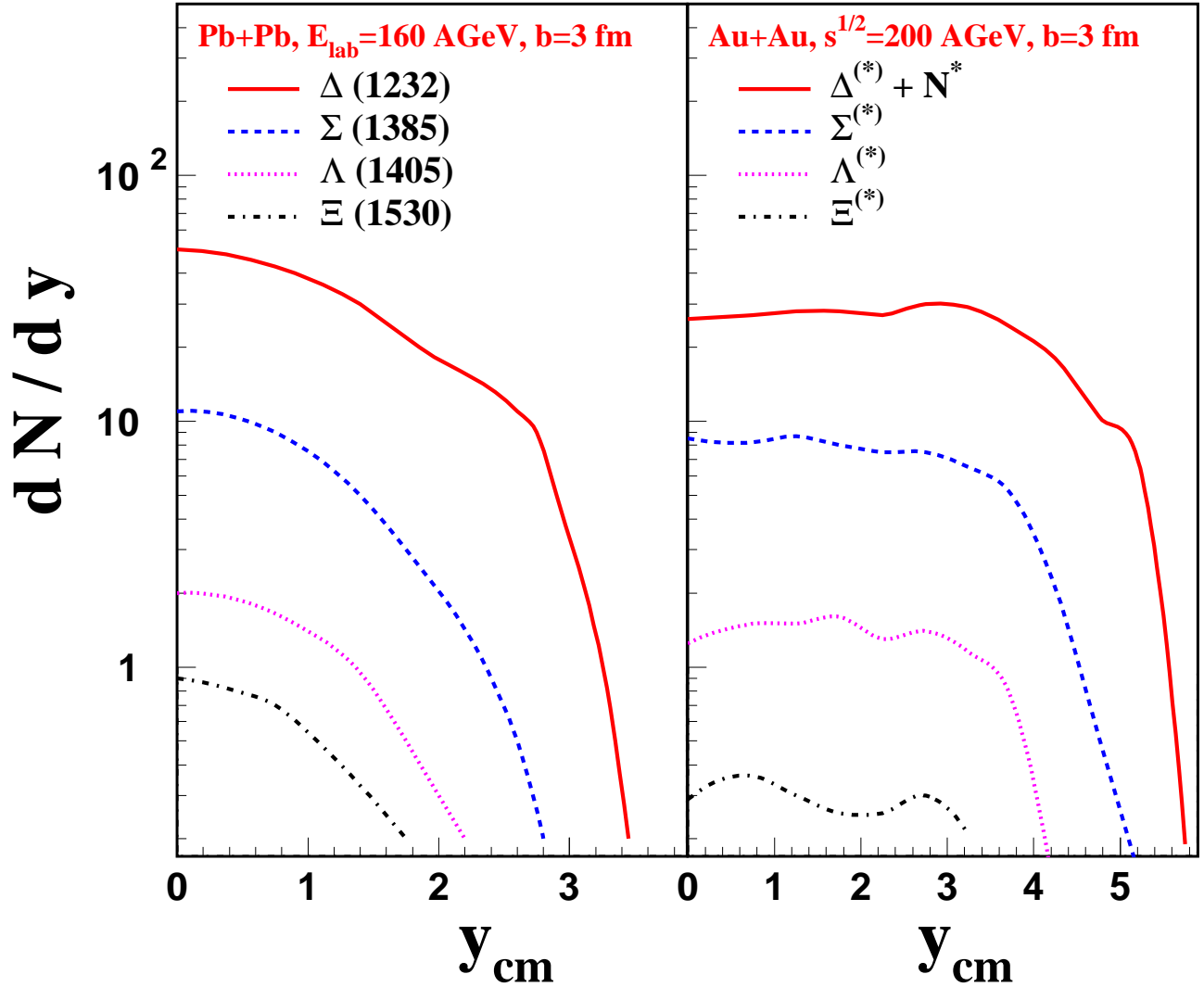}}
\caption{
The rapidity distributions of baryon resonances in Pb+Pb collisions
at $E_{lab} = 160${\it A\/}~GeV (left panel) and in Au+Au collisions
at $\sqrt{s} = 200${\it A\/}~GeV (right panel). Collisions are
calculated in the UrQMD with the impact parameter $b = 3$ fm.
}
\label{fig7}
\end{figure}

But can the formation of the resonance-abundant matter be traced
experimentally? To answer this question the rapidity distributions
of baryon resonances are plotted in Fig.~\ref{fig7} for central
($b = 3$ fm) Pb+Pb collisions at $E_{lab} = 160${\it A\/}~GeV and
Au+Au collisions at $\sqrt{s} = 200${\it A\/}~GeV, respectively. 
(The UrQMD predictions for other global observables at RHIC 
energies can be found in \cite{Blei00}).
Here only those resonances that decay into ground state hadrons,
i.e. no final state interactions, have been accounted for. 
The rapidity distributions of baryon resonances at SPS energies 
have a characteristic Gaussian-like shape, while the $d N/ d y$
distributions of those at RHIC energies are nearly flat in the
rapidity interval $|y| \leq 3.5$ resembling the Bjorken scaling 
picture of nuclear matter expansion at ultra-relativistic energies
\cite{Bjor83}. More than 80\% of the baryon non-strange resonances
at RHIC energies are still $\Delta$'s (1232).
One can see that the density of directly reconstructible baryon
resonances, especially $\Delta$'s and $\Lambda$'s + $\Sigma$'s, per 
unit rapidity at RHIC energies is quite high. Therefore, the
resonance-rich hadron matter produced in central Au+Au collisions
at $\sqrt{s} = 200${\it A\/}~GeV can be detected.

\section{Conclusions}
\label{sec5}

In summary, the microscopic transport  model UrQMD is applied to 
study the equation of state of the hot meson-dominated hadron-string
matter produced in the central  cell ($V = 125$ fm$^3$) of Au+Au
collisions at $\sqrt{s} = 200${\it A\/} GeV. After the restoration of 
the isotropy of pressure gradients, the hadron spectra in the cell 
are compared with those of the statistical model of an ideal hadron
gas. It is found that the expansion of matter in central collisions
proceeds with constant entropy per baryon ratio in the central 
cell, $S/A = s/\rho_{\rm B} \cong 150$. Since the $S/A$ ratios for
the central cell in A+A collisions, calculated at AGS and SPS
energies, are very close to the ratios extracted from the analysis
of the experimental data, the expected value of the entropy per
baryon ratio at RHIC lies within the range $150 \leq s/\rho_{\rm B} 
\leq 170$. The microscopic pressure in the cell is also close to the 
SM pressure. It shows a linear dependence on the energy density in 
the cell, $P = 0.15\, \varepsilon$, which is similar 
to the $P(\varepsilon)$ dependence in the central cell at SPS
energies \cite{lv99prc}. The obtained result is in accord with the
EOS $P \cong 0.2\, \varepsilon$, derived for an ideal gas of
hadrons and hadron resonances \cite{Shur73}. The temperature 
$T^{SM}$ in the cell at RHIC energies is shown to be nearly 
independent of the baryon chemical potential $\mu_{\rm B}$.

The further comparison of the energy spectra and yields of hadrons
with the SM predictions shows that the full thermal and chemical
equilibrium is not reached even at the late stage of the reaction.
This means that the times $t \approx 20$ fm/$c$ may be too short for
the relaxation process. Particularly, the deceleration of the 
relaxation to equilibrium is attributed to the creation of the
long-lived resonance-abundant matter. For instance, pions which are
frequently absorbed and produced in various inelastic processes,
including formation and decay of resonances, experience on average
only two elastic collisions per particle. This appears to be
insufficient to reach thermal equilibrium in the system of strongly
interacting particles. In turn, inelastic collisions try to restore 
chemical equilibrium in the cell. Non-equilibrium densities of 
strange and non-strange baryons, like $N$'s, $\Lambda$'s, $\Sigma$'s, 
etc, and their antiparticles are still lower than the equilibrium 
values. Amazingly, the yields of resonances are in accord with the SM 
values from the very early times $t \approx 5$ fm/$c$. Even the
abundances of pions are equalizing after $t \geq 15$ fm/$c$. However, 
this result should be taken with a grain of salt. The fitting 
temperature of the thermal model is higher than the inverse slope 
parameters of the energy spectra of particles in the cell, i.e., the 
temperature of the chemical freeze-out will be overestimated by the 
SM fit.

According to microscopic calculations, resonance-rich matter
survives until the thermal freeze-out when the contact between 
the hadrons is lost. It remains a challenging task to verify the
formation of long-lived resonance-abundant matter in heavy-ion
collisions at $\sqrt{s} = 200${\it A\/} GeV experimentally.

\acknowledgments
The authors profit from discussions with L. Csernai, 
M. Gorenstein, and E. Shuryak.
This work was supported by the Graduiertenkolleg f{\"u}r Theoretische
und Experimentelle Schwerionenphysik, Frankfurt--Giessen, the
Bundesministerium f{\"u}r Bildung und Forschung, the Gesellschaft
f{\"u}r Schwerionenforschung, Darmstadt, Deutsche
Forschungsgemeinschaft, and the Alexander von Humboldt-Stiftung, Bonn.
S.A.B was supported by the National Science Foundation, grant
PHY-00-70818.

\newpage

\newpage

\begin{table}
\caption{
The time evolution of the thermodynamic characteristics of hadronic
matter in the central cell of volume $V = 125$ fm$^3$ in central Au+Au
collisions at RHIC ($\sqrt{s}=200${\it A\/} GeV) energy. The 
temperature $T$, baryon chemical potential $\mu_{\rm B}$, strange 
chemical potential $\mu_{\rm S}$, pressure $P$, entropy density 
$s$, and entropy density per baryon density $s/\rho_{\rm B}$, are 
extracted from the statistical model of an ideal hadron gas, using the 
microscopically evaluated energy density $\varepsilon^{\rm cell}$, 
baryonic density $\rho_{\rm B}^{\rm cell}$, and strangeness density
$\rho_{\rm S}^{\rm cell}$ as input.
}
\begin{tabular}{@{}cccccccccc}
time & $\varepsilon^{\rm cell}$ & $\rho_{\rm B}^{\rm cell}$ &
$\rho_{\rm S}^{\rm cell}$ & $T$ & $\mu_{\rm B}$ & $\mu_{\rm S}$ &
$P$ & $s$ & $s/\rho_{\rm B}^{\rm cell}$ \\
fm/$c$ & MeV/fm$^3$ & fm$^{-3}$ & fm$^{-3}$ & MeV & MeV & MeV &
 MeV/fm$^3$ & fm$^{-3}$ &  \\
\tableline\tableline
 5 & 2330 & 0.093 & -0.0042 & 201 & 39.7 & 13.0 & 349 & 
       13.3 & 143 \\
 6 & 1705 & 0.071 & -0.0047 & 193 & 41.0 & 12.1 & 257 & 
       10.1 & 142 \\
 7 & 1319 & 0.059 & -0.0011 & 187 & 44.2 & 13.4 & 201 & 
        8.1 & 138 \\
 8 & 1031 & 0.045 &  0.0011 & 181 & 44.3 & 13.6 & 159 & 
        6.5 & 144 \\
 9 &  820 & 0.040 & -0.0044 & 176 & 47.2 & 10.4 & 128 & 
        5.4 & 135 \\
10 &  656 & 0.029 & -0.0022 & 171 & 44.1 &  9.8 & 104 & 
        4.4 & 152 \\
11 &  544 & 0.025 & -0.0044 & 167 & 47.4 & 11.2 &  87 & 
        3.8 & 149 \\
12 &  446 & 0.018 &  0.0089 & 163 & 42.6 & 10.6 &  72 & 
        3.2 & 176 \\
13 &  346 & 0.015 &  0.0040 & 158 & 50.4 & 15.6 &  57 & 
        2.6 & 165 \\
14 &  290 & 0.012 &  0.0024 & 154 & 48.0 & 12.9 &  49 & 
        2.2 & 180 \\
15 &  241 & 0.009 & -0.0044 & 150 & 40.7 &  6.5 &  41 & 
        1.9 & 214 \\
16 &  200 & 0.007 &  0.0022 & 147 & 37.8 &  6.6 &  35 & 
        1.6 & 250 \\
17 &  168 & 0.006 & -0.0031 & 143 & 42.7 & -1.1 &  30 & 
        1.4 & 218 \\
18 &  145 & 0.005 & -0.0028 & 140 & 43.4 & -1.9 &  26 &
        1.2 & 227 \\
\end{tabular}
\label{tab1}
\end{table}

\begin{table}
\caption{
The partial baryon densities of baryons and antibaryons,
$R_{\rm B/\bar{B}}$, given
by the microscopic model and obtained by the SM fit to the URQMD data,
within the time interval $5 \leq t \leq 19$ fm/$c$ in the central cell 
of Au+Au collisions at RHIC energies.
}
\begin{tabular}{@{}ccccc}
time & $R^{\rm mic}_{\rm B}$ & $ R^{\rm mic}_{\rm \bar{B}}$ 
     & $R^{\rm SM}_{\rm B} $ & $ R^{\rm SM}_{\rm \bar{B}}$      \\
fm/$c$ &  fm$^{-3}$  &  fm$^{-3}$  &  fm$^{-3}$  &  fm$^{-3}$   \\
\tableline\tableline
 5 & 0.183  & 0.090  & 0.332  & 0.239  \\
 6 & 0.135  & 0.064  & 0.236  & 0.165  \\
 7 & 0.106  & 0.047  & 0.178  & 0.119  \\
 8 & 0.082  & 0.036  & 0.134  & 0.088  \\
 9 & 0.065  & 0.025  & 0.105  & 0.065  \\
10 & 0.049  & 0.020  & 0.079  & 0.050  \\
11 & 0.041  & 0.016  & 0.064  & 0.039  \\
12 & 0.031  & 0.013  & 0.049  & 0.031  \\
13 & 0.025  & 0.010  & 0.036  & 0.021  \\
14 & 0.020  & 0.008  & 0.029  & 0.017  \\
15 & 0.016  & 0.007  & 0.022  & 0.013  \\
16 & 0.012  & 0.006  & 0.017  & 0.011  \\
17 & 0.010  & 0.004  & 0.014  & 0.008  \\
18 & 0.009  & 0.004  & 0.011  & 0.008  \\
19 & 0.008  & 0.003  & 0.010  & 0.005  \\
\end{tabular}
\label{tab2}
\end{table}

\begin{table}
\caption{
The partial strangeness densities, $R_{\rm S}^i$, of baryons, mesons, 
and their antiparticles, given
by the microscopic model and obtained by the SM fit to the URQMD data,
within the time interval $5 \leq t \leq 19$ fm/$c$ in the central cell 
of Au+Au collisions at RHIC energies.
}
\begin{tabular}{@{}ccccccccc}
time &
$(R_{\rm S}^{\rm B})^{\rm mic}$ &
$(R_{\rm \bar{S}}^{\rm \bar{B}})^{\rm mic}$ &
$(R_{\rm S}^{\rm B})^{\rm SM}$ &
$(R_{\rm \bar{S}}^{\rm \bar{B}})^{\rm SM}$ & 
$(R_{\rm \bar{S}}^{\rm M})^{\rm mic}$&$(R_{\rm S}^{\rm M})^{\rm mic}$ 
&$(R_{\rm \bar{S}}^{\rm M})^{\rm SM}$&$(R_{\rm S}^{\rm M})^{\rm SM}$\\ 
fm/$c$ & fm$^{-3}$ & fm$^{-3}$ & fm$^{-3}$ & fm$^{-3}$ & fm$^{-3}$ & 
fm$^{-3}$ & fm$^{-3}$ & fm$^{-3}$  \\
\tableline\tableline
 5 &-0.064 & 0.039 &-0.160 & 0.129 & 0.216 &-0.194 & 0.216 &-0.190  \\
 6 &-0.049 & 0.029 &-0.112 & 0.088 & 0.169 &-0.154 & 0.169 &-0.149  \\
 7 &-0.038 & 0.020 &-0.083 & 0.063 & 0.141 &-0.124 & 0.139 &-0.120  \\
 8 &-0.031 & 0.017 &-0.062 & 0.047 & 0.115 &-0.099 & 0.115 &-0.098  \\
 9 &-0.024 & 0.013 &-0.048 & 0.033 & 0.094 &-0.087 & 0.094 &-0.083  \\
10 &-0.019 & 0.010 &-0.036 & 0.025 & 0.077 &-0.071 & 0.078 &-0.070  \\
11 &-0.017 & 0.009 &-0.028 & 0.019 & 0.063 &-0.056 & 0.068 &-0.059  \\
12 &-0.014 & 0.007 &-0.021 & 0.015 & 0.052 &-0.045 & 0.058 &-0.050  \\
13 &-0.011 & 0.006 &-0.015 & 0.011 & 0.045 &-0.036 & 0.048 &-0.039  \\
14 &-0.009 & 0.005 &-0.012 & 0.008 & 0.038 &-0.031 & 0.041 &-0.035  \\
15 &-0.008 & 0.005 &-0.009 & 0.006 & 0.027 &-0.024 & 0.034 &-0.031  \\
16 &-0.006 & 0.004 &-0.007 & 0.005 & 0.023 &-0.020 & 0.029 &-0.026  \\
17 &-0.005 & 0.002 &-0.006 & 0.003 & 0.017 &-0.017 & 0.023 &-0.024  \\
18 &-0.005 & 0.002 &-0.005 & 0.003 & 0.014 &-0.015 & 0.020 &-0.021  \\
19 &-0.005 & 0.002 &-0.005 & 0.002 & 0.012 &-0.015 & 0.017 &-0.021  \\
\end{tabular}
\label{tab3}
\end{table}

\end{document}